\documentclass[aps,pre,final,10pt,showpacs,twocolumn,superscriptaddress,nofootinbib,flushbottom]{revtex4-1}%
\usepackage{amssymb,amsmath,amsfonts}
\usepackage{graphicx,color}
\usepackage[squaren]{SIunits}
\usepackage[english]{babel}
\usepackage{tabularx}

\newcolumntype{Y}{>{\centering\arraybackslash}X}

\newcommand{\dif}{\mathrm{d}}%
\newcommand{\norm}[1]{\lVert#1\rVert}%

\renewcommand{\vec}[1]{\boldsymbol{#1}}%
\renewcommand{\mathbf}[1]{\boldsymbol{#1}}%
\newcommand{\PP}{\hat{\mathcal{P}}}%

\begin{document}
\title{Pressure and Phase Equilibria in Interacting Active Brownian Spheres}

\author{Alexandre P. Solon}
\affiliation{Laboratoire, Mati\`ere et Syst\`emes Complexes, UMR 7057 CNRS/P7, Universit\'e Paris Diderot, 75205 Paris Cedex 13, France}

\author{Joakim Stenhammar}
\affiliation{SUPA, School of Physics and Astronomy, University of Edinburgh, Edinburgh EH9 3FD, United Kingdom}

\author{Raphael Wittkowski}
\affiliation{SUPA, School of Physics and Astronomy, University of Edinburgh, Edinburgh EH9 3FD, United Kingdom}

\author{Mehran Kardar}
\affiliation{Massachusetts Institute of Technology, Department of Physics, Cambridge, Massachusetts 02139, USA}

\author{Yariv Kafri}
\affiliation{Department of Physics, Technion, Haifa 32000, Israel}

\author{Michael E. Cates}
\affiliation{SUPA, School of Physics and Astronomy, University of Edinburgh, Edinburgh EH9 3FD, United Kingdom}

\author{Julien Tailleur}
\affiliation{Laboratoire, Mati\`ere et Syst\`emes Complexes, UMR 7057 CNRS/P7, Universit\'e Paris Diderot, 75205 Paris Cedex 13, France}

\date{\today}

\begin{abstract}
  {We derive a microscopic expression for the mechanical pressure
    $P$ in a system of spherical active Brownian particles at density
    $\rho$. Our exact result relates $P$, defined as the force per
    unit area on a bounding wall, to bulk correlation functions
    evaluated far away from the wall.} It shows that (i) $P(\rho)$ is
  a state function, independent of the particle-wall interaction; (ii)
  interactions contribute two terms to $P$, one encoding the slow-down
  that drives motility-induced phase separation, and the other a
  direct contribution well known for passive systems; (iii) {$P$
    is equal in coexisting phases.} We discuss the consequences of
  these results for the motility-induced phase separation of active
  Brownian particles, and show that the densities at coexistence
  \textit{do not} satisfy a Maxwell construction on $P$.
\end{abstract}


\maketitle

Much recent research addresses the statistical physics of active
matter, whose constituent particles show autonomous dissipative motion
(typically self-propulsion), sustained by an energy supply. Progress
has been made in understanding spontaneous flow \cite{RMP} and phase
equilibria in active matter~\cite{TC,CT,MIPS,fily,baskaran}, but as
yet there is no clear thermodynamic framework for these systems. Even
the definition of basic thermodynamic variables such as temperature
and pressure is problematic. While ``effective temperature'' is a
widely used concept outside equilibrium~\cite{Temperature}, the
discussion of pressure, $P$, in active matter has been neglected until
recently
\cite{Rosalind,Cristina,Brady,ModelB,Berthier,MalloryPRE2014,Niarxiv2014}.
At first sight, because $P$ can be defined mechanically as the force
per unit area on a confining wall, its computation as a statistical
average looks unproblematic. Remarkably though, it was recently shown
that for active matter the force on a wall can depend on details of
the wall-particle interaction so that $P$ is not, in general, a state
function \cite{Solon}.

Active particles are nonetheless clearly capable of exerting a
mechanical pressure $P$ on their containers. (When immersed in a
space-filling solvent, this becomes an {\em osmotic} pressure
\cite{Rosalind,Brady}.)  Less clear is how to calculate $P$; several
suggestions have been made \cite{Cristina,Brady,ModelB,Berthier} whose
inter-relations are, as yet, uncertain. Recall that for systems in
thermal equilibrium, the mechanical and thermodynamic definitions of
pressure (force per unit area on a confining wall, and
$-(\partial\mathcal{F}/\partial V)_{N}$ for $N$ particles in volume
$V$, with $\mathcal{F}$ the Helmholtz free energy) necessarily
coincide. Accordingly, various formulae for $P$ (involving, e.g., the
density distribution near a wall \cite{henderson}, or correlators in
the bulk \cite{allentildesley,Doi}) are always equivalent. This ceases
to be true, in general, for active particles \cite{ModelB,Solon}.

In this Letter we adopt the mechanical definition of $P$. We first
show analytically that $P$ is a state function, independent of the
wall-particle interaction, for one important and well-studied class of
systems: spherical active Brownian particles (ABPs) with isotropic
repulsions.  By definition, such ABPs undergo overdamped motion in
response to a force that combines an arbitrary pair interaction with
an external forcing term of constant magnitude along a body axis; this
axis rotates by angular diffusion. While not a perfect representation
of experiments (particularly in bulk fluids, where self-propulsion is
created internally and hydrodynamic torques arise \cite{Suzanne}),
ABPs have become the mainstay of recent simulation and theoretical
studies
\cite{fily,baskaran,CT,sten1,sten2,gompper,bechinger,speck}. They
provide a benchmark for the statistical physics of active matter, and
a simplified model for the experimental many-body dynamics of
autophoretic colloidal swimmers, or other active systems, coupled to a
momentum reservoir such as a supporting surface
\cite{sheffield,palacci,bocquet,herminghaus,bechinger,footroll}.  (We
comment below on the momentum-conserving case.)  By generating large
amounts of data in systems whose dynamics and interactions are
precisely known, ABP simulations are currently better placed than
experiments to answer fundamental issues concerning the physics of
active pressure, such as those raised in \cite{Cristina,Brady}.

Our key result exactly relates $P$ to bulk correlators, powerfully
generalizing familiar results for the passive case. {The pressure
  for ABPs is the sum of an ideal-gas contribution and a non-ideal one
  stemming from interactions. Crucially, the latter results from {\em
    two} contributions: one is a standard, `direct' term (the density
  of pairwise forces acting across a plane), which we call
  $P_\mathrm{D}$, while the other, `indirect' term, absent in the
  passive case, describes the reduction in momentum flux caused by
  collisional slowdown of the particles.  For short-ranged repulsions
  and high propulsive force, $P_\mathrm{D}$ becomes important only at
  high densities; the indirect term dominates at intermediate
  densities and is responsible for motility-induced phase separation
  (MIPS) \cite{TC,CT,MIPS}. The same calculation establishes that, for
  spherical ABPs (though not in general \cite{Solon}) $P$ must be
  equal in all coexisting phases.}

We further show that our ideal and indirect terms together form
exactly the `swim pressure', $P_{\mathrm{S}}(\rho)$ at density $\rho$,
previously defined via a force-moment integral in
\cite{Cristina,Brady}, and moreover that (in 2D) $P_{\mathrm{S}}$ is
simply $\rho v(0) v(\rho)/(2D_{\mathrm{r}})$, where $v(\rho)$ is the
mean propulsive speed of ABPs and $D_{\mathrm{r}}$ their rotational
diffusivity. We interpret this result, and show that (for
$P_\mathrm{D} = 0$) the mechanical instability, $\dif
P_{\mathrm{S}}/\dif\rho = 0$, coincides exactly with a diffusive one
previously found to cause MIPS among particles whose interaction
comprises a density-dependent swim speed $v(\rho)$
\cite{TC,CT,MIPS}. We briefly explain why this correspondence does not extend
to phase equilibria more generally, {{deferring a full account to a longer paper \cite{Longer}.}}

To calculate the pressure in interacting ABPs, we follow~\cite{Solon}
and consider the dynamics in the presence of an explicit, conservative
wall-particle force $\vec{F}_{\mathrm{w}}$. {For simplicity, we
  work in {2D}, {and} consider periodic boundary conditions in $y$
  and confining walls parallel to ${\vec e_y}=(0,1)$.} We start from the
standard Langevin dynamics of ABPs with bare speed $v_{0}$,
interparticle forces $\vec{F}$ and unit mobility
\cite{fily,baskaran,Henkes}:
\begin{equation}
\begin{split}
\dot{\vec{r}}_i &= v_{0}\vec{u}(\theta_i) + F_{\mathrm{w}}(x_i) \vec{e}_{x} + \sum_{j\neq i}
\vec{F}(\vec{r}_j-\vec{r}_i)+ \sqrt{2D_{\mathrm{t}}}\vec{\eta}_i \;, \\ 
\dot{\theta}_i &= \sqrt{2D_{\mathrm{r}}}\xi_i \;.
\end{split}\raisetag{6ex}\label{Lang1}%
\end{equation}
Here $\vec{r}_i(t) = (x_i,y_i)$ is the position, and $\theta_i(t)$ the
orientation, of particle $i$ at time $t$; $\vec{u}(\theta) =
(\cos(\theta),\sin(\theta))$; $F_{\mathrm{w}}=\norm{\boldsymbol{F}_{\mathrm{w}}}$ is a force acting along the
wall normal $\vec{e}_{x} = (1,0)$; $D_{\mathrm{t}}$ is the bare translational diffusivity;
 and $\vec{\eta}_i(t)$
and $\xi_i(t)$ are zero-mean unit-variance Gaussian white noises with
no correlations among particles.

Following standard procedures \cite{Dean,TC,CT,Farrell2012} this leads
to an equation for the fluctuating distribution function
$\hat{\psi}(\vec{r},\theta,t)$ whose zeroth, first, and second angular
harmonics are the fluctuating particle density $\hat{\rho}
=\int\hat{\psi}\,\dif\theta$; the $x$-polarization $\PP = \int
\hat{\psi} \cos(\theta) \,\dif\theta$; and $\hat{\mathcal{Q}} =
\int\hat{\psi} \cos(2\theta)\,\dif\theta$, which encodes nematic order
normal to the wall:
\begin{equation}
\begin{split}
&\dot{\hat{\psi}} = -\nabla\!\cdot\!\Big(\! \big(v_{0}\vec{u}(\theta) + F_{\mathrm{w}}(x)\vec{e}_{x} + \!\int \!\vec{F}(\vec{r}'-\vec{r}) \hat{\rho} (\vec{r}')\,\dif^2r'\big)\hat{\psi}\Big) \\
&\;+D_{\mathrm{r}}\partial^2_\theta\hat{\psi} + D_{\mathrm{t}}\nabla^2\hat{\psi} + \nabla\!\cdot\!\big(\sqrt{2D_{\mathrm{t}}\hat{\psi}}\vec{\eta}\big) + \partial_\theta\big(\sqrt{2D_{\mathrm{r}}\hat{\psi}}\xi\big) \;,
\end{split}\raisetag{5ex}
\label{Lang2}
\end{equation} 
{where $\vec{\eta}(\vec{r},t)$ and $\xi(\vec{r},t)$ are
  $\delta$-correlated, zero-mean, and unit-variance, Gaussian white
  noise fields.}  In steady-state, the noise-averages $\rho = \langle
\hat \rho \rangle$, $\mathcal{P} = \langle \hat{ \mathcal{P}}\rangle$,
and $\mathcal{Q} = \langle \hat{\mathcal{Q}}\rangle$ are, by
translational invariance, functions of $x$ only, as is the wall force
$F_{\mathrm{w}}(x)$ \cite{TI}. {Integrating \eqref{Lang2} over
  $\theta$, and then averaging over noise in steady state gives
  $\partial_{x}J = 0$, with $J$ the particle current. For any system
  with impermeable boundaries, {$J = 0$. Writing this out explicitly
    gives:}}
%
%
\begin{align}
0 &= v_{0} \mathcal{P} + F_{\mathrm{w}}\rho - D_{\mathrm{t}}\partial_x\rho + I_1(x) \;, \label{current}\\
I_1(x) &\equiv \int \! F_x(\vec{r}'-\vec{r}) \langle \hat{\rho} (\vec{r}')\hat{\rho} (\vec{r})\rangle \,\dif^2r' \;.
\label{I1}
\end{align}
Applying the same procedure to the first angular harmonic gives
\begin{align}
D_{\mathrm{r}}\mathcal{P} &= -\partial_x \! \left[\frac{v_{0}}{2}(\rho + \mathcal{Q}) + F_{\mathrm{w}}\mathcal{P} - D_{\mathrm{t}}\partial_x \mathcal{P} + I_2(x)\right] ,\quad \label{K} \raisetag{4ex} \\
I_2(x) &\equiv 
\int \! F_x(\vec{r}'-\vec{r}) \langle \hat{\rho} (\vec{r}')\PP(\vec{r})\rangle \,\dif^2r' \;.
\label{I2}
\end{align}
Note that the integrals $I_1$ and $I_2$ defined in \eqref{I1} and
\eqref{I2} are, by translational invariance, functions only of $x$.

{The mechanical pressure on the wall is the spatial integral of
  the force density exerted upon {it by} the particles.  The
  wall force obeys $F_{\mathrm{w}} = -\partial_x U_{\mathrm{w}}$ where
  {an origin is chosen so that}
  $U_{\mathrm{w}}$ is {non-zero only for
  $x>0$. The wall is confining, i.e. $F_{\mathrm{w}}\rho \to 0$ for
  $x\gg 0$, whereas $x = \Lambda \ll 0$ denotes any plane in the bulk of the fluid, far from the wall.} By Newton's third law, the pressure is then}
\begin{equation}
P = -\int_{\Lambda}^{\infty} \!\!\!\!\!\! F_{\mathrm{w}}(x)\rho(x)\,\dif x \;,
\label{Pm}
\end{equation}
{In} \eqref{Pm} we now use
\eqref{current} to set $-F_{\mathrm{w}}\rho = v_{0}\mathcal{P}
-D_{\mathrm{t}}\partial_x\rho+I_1${:}
\begin{equation}
P = v_{0}\int_\Lambda^\infty \!\!\!\!\!\! \mathcal{P}(x)\,\dif x + D_{\mathrm{t}}\rho(\Lambda) 
+\int_\Lambda^\infty \!\!\!\!\!\! I_1(x)\, \dif x \;.
\label{seven}
\end{equation}
We next use \eqref{K}, in which $\mathcal{P}$ and $\mathcal{Q}$ vanish
in the bulk and all terms vanish at infinity, to evaluate $\int
\mathcal{P}\,\dif x$, giving:
\begin{equation}
P=\frac{v_{0}}{D_{\mathrm{r}}}\left(\frac{v_{0}}{2}\rho(\Lambda) + I_2(\Lambda)\right)+D_{\mathrm{t}}\rho(\Lambda) + \int_\Lambda^\infty \!\!\!\!\!\! I_1(x)\, \dif x \;.
\label{Pmm}
\end{equation}
Using Newton's third law, the final integral in \eqref{Pmm} takes a
familiar form, describing the density of pair forces acting across
some plane through the bulk (far from any wall):
\begin{equation}
\int_{x>\Lambda} \!\!\!\!\!\!\!\! \dif x \int_{x'<\Lambda} \!\!\!\!\!\!\!\! \dif^2r'\;
F_x(\vec{r}'-\vec{r}) \langle \hat{\rho}(\vec{r}')\hat{\rho}(\vec{r})\rangle \equiv P_{\mathrm{D}} \;.
\label{PD}
\end{equation}
Thus in the passive limit ($v_{0} = 0$) we recover in $P_{\mathrm{D}}$
the standard interaction part in the pressure \cite{Doi}. We call
$P_{\mathrm{D}}$ the ``direct" contribution; it is affected by
activity only through changes to the correlator. Activity also enters
(via $v_{0}$) the well-known ideal pressure term
\cite{MalloryPRE2014,Brady,Cristina,Solon}:
\begin{equation}
P_0 \equiv \left(D_{\mathrm{t}}+\frac{v_{0}^2}{2D_{\mathrm{r}}}\right)\rho(\Lambda) \;.
\label{P0}
\end{equation}
Having set friction to unity in \eqref{Lang1}, $D_{\mathrm{t}} =
k_{\mathrm{B}}T$, so that within $P_0$ (only) activity looks like a
temperature shift.

Most strikingly, activity in combination with interactions also brings
an ``indirect" pressure contribution
\begin{equation}
P_{\mathrm{I}} \equiv \frac{v_{0}}{D_{\mathrm{r}}}I_2(\Lambda)
\label{PI}
\end{equation} 
with no passive counterpart. Here $I_2(\Lambda)$ is again a
wall-independent quantity, evaluated on {\em any} bulk plane
$x=\Lambda\ll 0$. We discuss this term further below.

Our exact result for mechanical pressure is finally
\begin{equation}
P=P_0+P_{\mathrm{I}}+P_{\mathrm{D}} 
\label{total}
\end{equation}
with these three terms defined by \eqref{P0}, \eqref{PI}, and
\eqref{PD}, respectively. $P$ is thus for interacting ABPs a state
function, calculable solely from bulk correlations and independent of
the particle-wall force $F_{\mathrm{w}}(x)$. Because the same boundary
force can be calculated using {\em any} bulk plane $x=\Lambda$, it
follows that, should the system undergo phase separation, $P$ is the
same in all coexisting phases \cite{TI}. This proves for ABPs an
assumption that, while plausible \cite{Brady,Brady2}, is not obvious,
and indeed can fail for particles interacting via a density-dependent
swim speed rather than direct interparticle forces \cite{Solon}.

Notably, although ABPs exchange momentum with a reservoir,
(\ref{Lang1}) also describes particles swimming through a
momentum-conserving bulk fluid, in an approximation where
inter-particle and particle-wall hydrodynamic interactions are both
neglected. So long as the wall interacts {\em solely} with the
swimmers, our results above continue to apply to what is now the {\em
  osmotic} pressure.

The physics of the indirect contribution $P_{\mathrm{I}}$ is that
interactions between ABPs reduce their motility as the density
increases. The ideal pressure term $P_0$ normally represents the flux
of momentum through a bulk plane carried by particles that {\em move}
across it (as opposed to those that {\em interact} across it)
\cite{allentildesley}. In our overdamped system one should replace in
the preceding sentence `momentum' with `propulsive force' (plus a
random force associated with $D_{\mathrm{t}}$). Per particle, the
propulsive force is density-independent, but the rate of crossing the
plane is not. Accordingly we expect the factor $v_{0}^2$ in \eqref{P0}
to be modified by interactions, with one factor $v_{0}$ (force or
momentum) unaltered, but the other (speed) replaced by a
density-dependent contribution $v(\rho) \le v_{0}$:
\begin{equation} 
P_0+P_{\mathrm{I}} =
\left(D_{\mathrm{t}}+\frac{v_{0}v(\rho)}{2D_{\mathrm{r}}}\right)\rho \;.
\label{P0I}
\end{equation}
This requires the mean particle speed to obey
\begin{equation}
v(\rho) = v_{0}+2I_2/\rho \;.
\label{speed}
\end{equation}
{Remarkably, \eqref{P0I} and \eqref{speed} are {\em exact} results,
where \eqref{speed} is found from the mean speed of particle $i$ in
bulk}, $v = v_{0} +\langle \vec{u}(\theta_i)\!\cdot\!\sum_{j\neq
  i}\vec{F}(\vec{r}_{j}-\vec{r}_{i})\rangle$. To see why this average
involves $I_2$, note that the system is isotropic in bulk, so $x$ and
$y$ can be interchanged in $I_2(x)$, and that $\cos(\theta) \equiv
\vec{u}\!\cdot\!\vec{e}_{x}$. Relation \eqref{I2} then links $v$ to
$I_2$ via the $\langle\hat{\rho} \PP \rangle$ correlator, which
describes the imbalance of forces acting on an ABP from neighbors in
front and behind.

\begin{figure}
\includegraphics[width=0.40\textwidth]{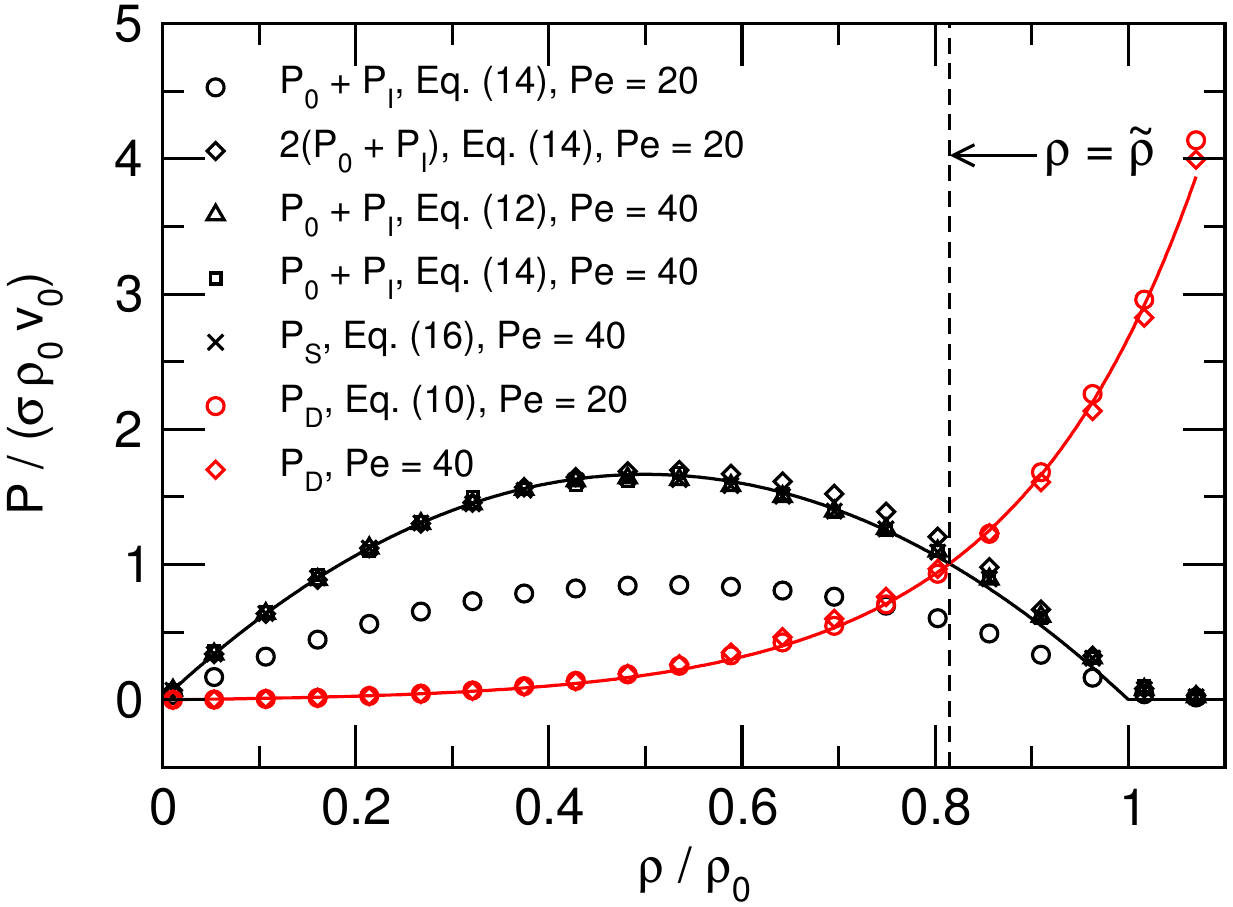}
\caption{{Numerical measurements of $P_0+P_{\mathrm{I}}$,
    $P_\mathrm{S}$, and $P_\mathrm{D}$ in single-phase ABP simulations
    at P\'eclet number Pe $\equiv 3v_{0}/(D_{\mathrm{r}}\sigma)$ = 40,
    where $\sigma$ is the particle diameter. Expressions~\eqref{PI},
    \eqref{P0I}, and \eqref{PS} for $P_0+P_\mathrm{I}$ and
    $P_\mathrm{S}$ show perfect agreement.} Also shown is data for Pe
  = 20, unscaled and rescaled by factor 2. This confirms that
  $P_{\mathrm{S}} = P_0+P_{\mathrm{I}}$ is almost linear in Pe; small
  deviations arise from Pe-dependence of the correlators. In
  {red} is $P_{\mathrm{D}}$ for Pe $= 20,40$, with no
  rescaling. {Pe was varied using $D_\mathrm{r}$, at fixed
    $v_{0}$ and with $D_\mathrm{t}=D_\mathrm{r}\sigma^2/3$}. Solid
  lines are fits to piecewise parabolic ($P_{\mathrm{S}}$) and
  exponential ($P_{\mathrm{D}}$) functions used in the semi-empirical
  equation of state. {$\rho_0$ is a near-close-packed density at which
  $v(\rho)$ vanishes and $\tilde \rho$ is the treshold density
  above which $P_\mathrm{D}>P_\mathrm{S}$. See~\cite{SI} for
    details}.
  \label{fig:pressures}}
\end{figure}

{Furthermore, the self-propulsive term in \eqref{P0I} is exactly
  the `swim pressure' $P_{\mathrm{S}}$ of \cite{Brady,Cristina}:}
\begin{equation}
\frac{v_{0}v(\rho)}{2D_{\mathrm{r}}}\rho = P_{\mathrm{S}} \equiv \frac{\rho}{2}\langle \vec{r}\!\cdot\!\vec{F}^{\mathrm{a}}\rangle 
\label{PS}
\end{equation}
with $\vec{F}^{\mathrm{a}} = v_{0}\vec{u}$ a particle's propulsive
force and $\vec{r}$ its position. (The particle mobility
$v_0/F^{\mathrm{a}} = 1$ in our units.) {The equivalence of
  \eqref{PI}, \eqref{P0I}, and \eqref{PS} is proven analytically
  in~\cite{SI} and confirmed numerically in Fig.~\ref{fig:pressures}
  for ABP simulations performed as in~\cite{sten1,sten2}.}

Thus for $D_{\mathrm{t}} = 0$, (\ref{total}) may alternatively be
rewritten as $P = P_{\mathrm{S}}+P_{\mathrm{D}}$
\cite{Brady,Cristina}. Together, our results confirm that
$P_{\mathrm{S}}$, defined in bulk via \eqref{PS}, determines (with
$P_{\mathrm{D}}$) the force acting on a confining wall. This was
checked numerically in \cite{Cristina} but is not automatic
\cite{Solon}. Moreover, our work gives via \eqref{P0I} an exact
kinetic expression for $P_{\mathrm{S}}$ with a clear and simple
physical interpretation in terms of the transport of propulsive
forces. This illuminates the nature of the swim pressure
$P_{\mathrm{S}}$ and extends to finite $\rho$ the limiting result $P_\mathrm{S}
= P_0$ \cite{Cristina,Brady}.

The connections made above are our central findings; they extend
statistical thermodynamics concepts from equilibrium far into ABP
physics.  Before concluding, we ask how far these ideas extend to
phase equilibria.

In the following we ignore for simplicity the $D_{\mathrm{t}}$ term
(negligible in most cases~\cite{fily,sten1,CT,Henkes}). Then, assuming
short-range repulsions, we have {$P_{\mathrm{S}} = \rho
  v_{0}v(\rho)/(2D_{\mathrm{r}})$}, with $v(\rho) \simeq
v_{0}(1-\rho/\rho_0)$ and $\rho_0$ a near-close-packed density
\cite{fily,baskaran,sten1}. $P_{\mathrm{D}}$ should scale as
$\sigma\rho v_{0} \mathcal{S}(\rho/\rho_0)$, where $\sigma$ is the
particle diameter and the function $\mathcal{S}$ diverges at close
packing; here the factor $v_{0}$ is because propulsive forces oppose
repulsive ones, setting their scale
\cite{Brady}. {Figure~\ref{fig:pressures} shows {that both}
  the approximate expression {for} $P_{\mathrm S}$ (with a fitted
  $\rho_0\simeq 1.19$ roughly independent of Pe){,} and the scaling of
  $P_{\mathrm{D}}${,} hold remarkably well.}  Defining a threshold
value $\tilde{\rho}$ by $P_{\mathrm{S}}(\tilde{\rho}) =
P_{\mathrm{D}}(\tilde{\rho})$ (see Fig.~\ref{fig:pressures}), it
follows that at large enough P\'eclet number, Pe
$=3v_{0}/(D_{\mathrm{r}}\sigma)$, $P_{\mathrm{S}}$ dominates
completely for $\rho < \tilde{\rho}$, with $P_{\mathrm{D}}$ serving
{\em only} to prevent the density from moving above the $\tilde \rho$
cutoff.  When $\rho<\tilde{\rho}$, $P_{\mathrm{D}}$ is negligible; the
criterion $P_{\mathrm{S}}'(\rho) < 0$, used in \cite{Brady,Brady2} to
identify a mechanical instability, is then via \eqref{PS} {\em
  identical} to the spinodal criterion $(\rho v)'< 0$ used to predict
MIPS in systems whose sole physics is a density-dependent speed
$v(\rho)$ \cite{TC,CT}. Thus, for ABPs at large Pe, the mechanical
theory reproduces one result of a long-established mapping between
MIPS and equilibrium colloids with attractive forces \cite{TC,CT}.

We next address the binodal densities of coexisting phases. According
to \cite{TC,CT}, particles with speed $v(\rho)$ admit an effective
bulk free-energy density $f(\rho) =
k_{\mathrm{B}}T\left[\rho(\ln\rho-1) + \int^\rho_{0} \ln v(u) \,\dif
  u\right]$.  (Interestingly, the equality of $P$ in coexisting phases
is equivalent at high Pe and $\rho<\tilde\rho$ to the equality of
$k_\mathrm{B}T\log(\rho v)$, which is the chemical potential in this
`thermodynamic' theory \cite{TC,MIPS}.)  The binodals are then found
using a common tangent construction (CTC, i.e., global minimization)
on $f$, or equivalently an equal-area Maxwell construction (MC) on an
effective \textit{thermodynamic} pressure $P_f = \rho f'-f$, which
differs from $P$ \cite{ModelB}.  Formally, $f$ is a local
approximation to a large-deviation functional \cite{LDF}, whose
nonlocal terms can (in contrast to equilibrium systems) alter the
CTC or MC~\cite{sten1,ModelB}; we return to this issue below.

An appealing alternative is to apply the MC to the mechanical pressure
$P$ itself; this was, in different language, proposed in
\cite{Brady2}. {(The equivalence will be detailed
  in~\cite{Longer}.)} It amounts to constructing an effective
free-energy density $f_P(\rho)\neq f$, defined via $P = \rho
f_P'-f_P$, and using the CTC on $f_P$. However, $f_P$ has no clear
link to any large deviation functional \cite{LDF}; and since it
differs from $f$, these approaches {\em generically predict different
  binodals}.

\begin{figure}
\includegraphics[width=0.46\textwidth]{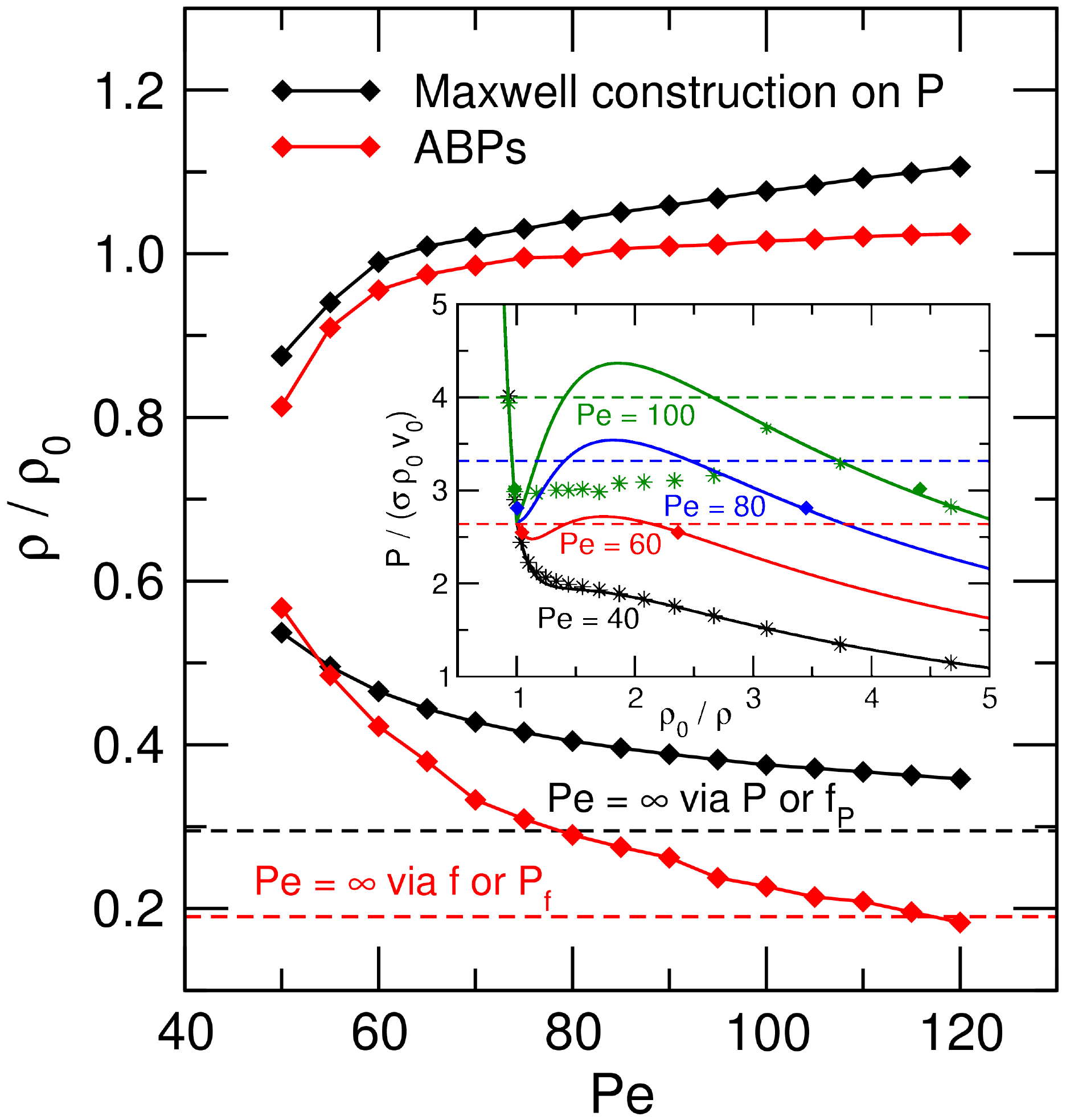}
\caption{Simulated coexistence curves (binodals) for ABPs {(red)}, and those
calculated via the Maxwell construction {(black)} on the mechanical
pressure $P$ {using the semi-empirical equation of state for $P_\mathrm{S}$ and $P_\mathrm{D}$ fitted from Fig.~\ref{fig:pressures}.}  Dashed lines: predicted high Pe
asymptotes for the binodals calculated via $f$ or $P_f$ (lower), and
calculated via $P$ or $f_P$ (upper). Inset: measured binodal
pressures and densities (diamonds) fall on the equation-of-state
curves but do not match the MC values (horizontal dashed lines). Stars
show the $P(\rho)$ relation across the full density range 
from simulations at Pe = 40 and Pe = 100. The latter includes two metastable
states at low density (high $\rho_0/\rho$) that are yet to phase separate.
\label{fig:PD}}
\end{figure}

To confirm this, we turn to the large Pe limit; here, for ABPs with
$v(\rho) = v_{0}(1-\rho/\rho_0)$ and $\tilde{\rho} = \rho_0$, we can
explicitly construct $f(\rho)$ (and hence $P_f(\rho)$) alongside
$P(\rho)$ (and hence $f_P(\rho)$), using our hard-cutoff approximation
(i.e., a constraint $\rho < \tilde\rho$). All four functions are
plotted in \cite{SI}; the two distinct routes indeed predict different
binodals at high Pe (see Fig.\ \ref{fig:PD}) \cite{softfoot}.  Each
approach suffers its own limitations.  That via $f$ (or $P_f$) appears
more accurate, but {neglects non-local terms} that can alter the
binodals: although $f'(\rho)$ remains equal in coexisting phases,
$P_f$ is not equal once those terms are included
\cite{ModelB}.  The most serious drawback of this approach,
currently, is that it cannot address finite Pe, where $P_{\mathrm{D}}$
no longer creates a sharp cutoff. Meanwhile the `mechanical' {route captures} the equality of $P$ in coexisting phases but
unjustifiably {assumes the MC on $P$},
asserting in effect that $f_P$, and not $f$, is the effective free
energy \cite{LDF}. Nonlocal corrections \cite{olmsted} are again
neglected.

At finite Pe where the crossover at $\tilde{\rho}$ is soft,
\eqref{total} shows how $P_{\mathrm{I}}$ and $P_{\mathrm{D}}$ compete,
giving Pe-dependent binodals (see Fig.\ \ref{fig:PD}). To test the
predictions of the mechanical approach {(equivalent to \cite{Brady2})}, we set
$P_{\mathrm{D}} = \sigma\rho v_{0}\mathcal{S}(\rho/\rho_0)$ as above,
finding the function $\mathcal{S}$ by numerics on single-phase systems
at modest Pe (see Fig.\ \ref{fig:pressures}). Adding this to
$P_{\mathrm{S}}$ (assuming $P_{\mathrm{S}}\propto$ Pe scaling) gives
$P = P(\rho,$Pe$)$. At each Pe the binodal pressures and densities do
lie on this equation of state, validating its semi-empirical form; but
they do not obey the Maxwell construction on $P$, which must therefore
be rejected (see Fig.\ \ref{fig:PD}, inset).
We conclude that, despite our work and that of \cite{Brady2}, no
complete theory of phase equilibria in ABPs yet exists.

In summary, we have given in \eqref{PD}-\eqref{total} an exact
expression for the mechanical pressure $P$ of active Brownian spheres.
This relates $P$ directly to bulk correlation functions and shows it
to be a state function, independent of the wall interaction, something
not true for all active systems \cite{Solon}. As well as an ideal term
$P_0$, and a direct interaction term $P_{\mathrm{D}}$, there is an
indirect term $P_{\mathrm{I}}$ caused by collisional slowing down of
propulsion. We established an exact link between $P_0 +
P_{\mathrm{I}}$ and the so called `swim pressure' \cite{Brady},
allowing a clearer interpretation of that quantity.
We showed that when MIPS arises in the regime of high Pe $=
3v_{0}/(D_{\mathrm{r}}\sigma)$, the mechanical ($P'<0$ \cite{Brady})
and diffusive ($f''<0$ \cite{TC,CT}) instabilities coincide.
That equivalence does not extend to the calculation of coexistence
curves, for reasons we have explained. {For simplicity we have worked
in 2D; generalization of our results to 3D is
straightforward~\cite{Longer} but notationally cumbersome.}

The established description of MIPS as a diffusive instability
\cite{TC,CT,sten1,ModelB} is fully appropriate in systems whose
particles are `programmed' to change their dynamics at high density
(e.g., via bacterial quorum sensing \cite{quorum,quorum2}), but it is
not yet clear whether the same theory, or one based primarily on the
mechanical pressure $P$, is better founded for finite-Pe phase
equilibria in ABPs whose slowdown is collisional.
Meanwhile, our exact results for $P$ in these systems add
significantly to our growing understanding of how statistical
thermodynamic concepts can, and cannot, be applied in active
materials.

{\it Acknowledgments:} We thank Rosalind Allen, John Brady, Cristina
Marchetti, and Xingbo Yang for seminal discussions. This work was
funded in part by EPSRC Grant EP/J007404. JS is supported by the
Swedish Research Council (350-2012-274), RW is supported through a
Postdoctoral Research Fellowship (WI 4170/1-2) from the German
Research Foundation (DFG), YK is supported by the I-CORE Program of
the Planning and Budgeting Committee of the Israel Science Foundation,
and MEC is supported by the Royal Society. APS and JT are supported by
ANR project BACTTERNS. APS, JS, MK, MEC, and JT thank the KITP at the
University of California, Santa Barbara, where they were supported
through National Science Foundation Grant NSF PHY11-25925.


\clearpage
\newpage
\begin{center}
  {\LARGE Supplementary Material}
\end{center}

\section{Proof of relation $\boldsymbol{P_0+P_{\mathrm{I}} = P_{\mathrm{S}}}$} 
We prove here (setting $D_{\mathrm{t}}= 0$ for simplicity, and working
in $d=2$ dimensions) that the sum of the ideal pressure $P_0 = \rho
v_0^2/(2D_{\mathrm{r}})$ and the indirect interaction pressure
$P_{\mathrm{I}}=v_0I_2/D_{\mathrm{r}}$ is the swim pressure
$P_{\mathrm{S}} = \rho \langle\vec{r}\!\cdot\!
\vec{F}^{\mathrm{a}}\rangle/2$, where the self-propulsion force
$\vec{F}^{\mathrm{a}}=v_0\vec{u}$ was defined in (16) in the main
text.  (As in the main text, we set the particle mobility
$v_0/F^\mathrm{a} = 1$ in this section, where
$F^\mathrm{a}\equiv\norm{\vec{F}^\mathrm{a}}$.)  In proving the
required result, we also establish that $P_{\mathrm{S}} =\rho v_0
v(\rho)/(2D_{\mathrm{r}})$, and hence that $v(\rho) = v_0+2I_2/\rho$.

We start from (see (6) in the main text)
\begin{equation}
I_2 = \int \! F_x(\vec{r}'-\vec{r}) \langle \hat{\rho}(\vec{r}')\PP (\vec{r})\rangle\,\dif^{2}r' 
\label{I2}%
\end{equation}
and use\footnote{Equations \eqref{rhom} and \eqref{Pm} follow from $\hat{\psi}(\vec{r},\theta)= \sum_i\delta(\vec{r}-\vec{r}_i)\delta(\theta-\theta_i)$ and the definitions of $\hat{\rho}(\vec{r})$ and $\PP(\vec{r})$.}
\begin{align}
\hat{\rho}(\vec{r})&=\sum_i \delta(\vec{r}-\vec{r}_i) \;,\label{rhom}\\
\PP(\vec{r})&=\sum_i \cos(\theta_i) \delta(\vec{r}-\vec{r}_i) \label{Pm}
\end{align}
as well as the fact that the system is isotropic to rewrite \eqref{I2}
in the form
\begin{equation}
I_2 = \frac{1}{L_xL_y} \bigg\langle \sum_{i,j\neq i} F_x(\vec{r}_j-\vec{r}_i)\cos(\theta_i) \bigg\rangle \,.
\label{one}%
\end{equation}
We now take the thermodynamic {limit:} $L_x = L_y =
\sqrt{A}\to\infty$.  Since the system is isotropic, a similar
expression can be written interchanging $x$ and $y$, noting that
$\cos(\theta_i) = \vec{u}_i\!\cdot\! \vec{e}_x$ with
$\vec{u}_i=(\cos(\theta_i),\sin(\theta_i))$ and
$\vec{e}_x=(1,0)$. Averaging the two results gives
\begin{equation}
P_{\mathrm{I}} = \frac{v_0}{2D_{\mathrm{r}} A} \bigg\langle \sum_{i,j\neq i} \vec{F}(\vec{r}_j-\vec{r}_i)\!\cdot\!  \vec{u}_i \bigg\rangle \,.
\label{two}
\end{equation}  
We may also write, using the fact that $\vec{u}\!\cdot\! \vec{u} = 1$,
\begin{equation}
P_0 = \frac{\rho v_0^2}{2D_{\mathrm{r}}} = \frac{v_0}{2D_{\mathrm{r}}A}\sum_{i}v_0\langle \vec{u}_i \!\cdot\!  \vec{u}_i\rangle \;.
\end{equation}  
Hence, we obtain
\begin{equation}
P_0+P_{\mathrm{I}} = \frac{v_0}{2D_{\mathrm{r}} A} \bigg\langle\sum_{i}\Big( v_0\vec{u}_i+ \sum_{j\neq i}\vec{F}(\vec{r}_j-\vec{r}_i) \Big) \!\cdot\! \vec{u}_i \bigg\rangle \,.
\label{four}
\end{equation}  

From the Langevin equation (1) in the main text, applied in bulk where
the wall force $F_{\mathrm{w}}$ vanishes, and setting $D_{\mathrm{t}}
= 0$, we have that the term $v_0\vec{u}_i+ \sum_{j\neq
  i}\vec{F}(\vec{r}_j-\vec{r}_i)$ in \eqref{four} is the instantaneous
particle velocity $\dot{\vec{r}}_i$:
\begin{equation}
P_0+P_{\mathrm{I}} = \frac{v_0}{2D_{\mathrm{r}} A}\Big\langle\sum_{i}\dot{\vec{r}}_i\!\cdot\! \vec{u}_i\Big\rangle \,.
\end{equation}  
If we redefine $\langle\,\!\cdot\! \,\rangle$ to include an average
over the particle index, this may be written\begin{equation}
  P_0+P_{\mathrm{I}} = \frac{\rho
    v_0}{2D_{\mathrm{r}}}\langle\dot{\vec{r}}\!\cdot\! \vec{u}\rangle
  = \frac{\rho v_0}{2D_{\mathrm{r}}} v(\rho) \;.
\label{six}
\end{equation}  
Here, the second equality follows from the definition of
$v(\rho)\equiv\langle\dot{\vec{r}}\!\cdot\! \vec{u}\rangle$ as the
average speed of a particle along its propulsive direction (in a bulk
system at density $\rho$).

Meanwhile, $P_{\mathrm{S}}$ is defined via (16) in the main text
(setting $d=2$ there) as an equal-time average
\begin{equation}
P_{\mathrm{S}} = \frac{\rho}{2}\langle \vec{r}\!\cdot\!  \vec{F}^{\mathrm{a}} \rangle = 
\frac{\rho v_0}{2}\langle \vec{r}\!\cdot\! \vec{u}\rangle  \;.
\label{bradyp}
\end{equation}

We rewrite $\vec{r}(t) = \vec{r}(-\infty)
+\int_{-\infty}^{t}\!\dot{\vec{r}}(t')\,\dif t'$, and use time
stationarity and the fact that $\langle \vec{r}(-\infty)\!\cdot\!
\vec{u}(t)\rangle = 0$ to obtain
\begin{equation}
\langle \vec{r}(t)\!\cdot\! \vec{u}(t)\rangle = \int_0^\infty \!\!\!\!\! \langle \dot{\vec{r}}(0)\!\cdot\! \vec{u}(t')\rangle\, \dif t'  \;.
\label{integral}
\end{equation}
Next, we use the fact that the angular dynamics of $\vec{u}$ are
{autonomous: the rotational diffusion of one particle} is unaffected
by the location and orientation of any other particle. {Then
  $\dot{\vec{r}}(0)$ and $\vec{u}(t')$ are correlated, but only
  because each} is separately correlated with $\vec{u}(0)$.  That
separability allows us to write
\begin{equation}
\! \langle \dot{\vec{r}}(0)\!\cdot\! \vec{u}(t')\rangle = 
\frac{1}{2\pi}\! \int \langle \dot{\vec{r}}(0)|\vec{u}(0)\rangle \!\cdot\!  \langle \vec{u}(t')|\vec{u}(0)\rangle \, \dif \theta(0) \;, \! \label{conditionals}
\end{equation}
where the integration is over the bulk steady state orientations
$\theta(0)=\arccos(u_x(0))$ with uniform probability density
$1/(2\pi)$, and $\langle \vec{X}|\vec{Y}\rangle$ denotes the
conditional average of $\vec{X}$ {given $\vec{Y}$}. The first
conditional average in (\ref{conditionals}) obeys
\begin{equation}
\langle \dot{\vec{r}}(0)|\vec{u}(0)\rangle = v(\rho) \vec{u}(0) \;,
\label{cond1}
\end{equation}
which follows from the definition of $v(\rho)$ [see \eqref{six}] and
the fact that the mean velocity of a particle must point along its
axis $\vec{u}$, given isotropy of the bulk system. The second
conditional average in (\ref{conditionals}) is found from the
autonomous rotational dynamics as
\begin{equation}
\langle \vec{u}(t')|\vec{u}(0)\rangle = \vec{u}(0) \exp(-D_{\mathrm{r}} t') \;, 
\label{cond2}
\end{equation}
which (again given isotropy) is implied by the familiar decay of
angular correlations $\langle \vec{u}(t')\!\cdot\! \vec{u}(0)\rangle =
\exp(-D_{\mathrm{r}} t')$. It follows from \eqref{cond1} and
\eqref{cond2} that the product of the conditional averages in
\eqref{conditionals} is $v(\rho)\exp(-D_{\mathrm{r}} t')$, which is
independent of $\vec{u}(0)$ as befits an isotropic system. This gives
finally, upon performing the time integral in \eqref{integral},
\begin{equation} 
\langle \vec{r}\!\cdot\! \vec{u}\rangle = \frac{v(\rho)}{D_{\mathrm{r}}} \label{ito}\;,
\end{equation}
thus completing the proof that $P_{\mathrm{S}}$ defined by
\eqref{bradyp} is exactly equal to $P_0+P_{\mathrm{I}}$ as given by
\eqref{six}. Note that (\ref{ito}) can also be proved directly,
avoiding the use of conditional averages, by a route involving
It{\={o} calculus {\cite{Longer}}.

Having proved in \eqref{six} that (with $P_0=\rho v_0^2/(2D_{\mathrm{r}})$) the indirect pressure $P_{\mathrm{I}} = v_0I_2/D_{\mathrm{r}}$ obeys
\begin{equation}
P_{\mathrm{I}} = \frac{\rho v_0}{2D_{\mathrm{r}}}(v(\rho)-v_0) \;,
\end{equation}
it follows, as stated in the main text, that
\begin{equation}
v(\rho) = v_0+2I_2/\rho \;.
\end{equation}
We know from ABP simulations \cite{Stenhammar-softmatter} that, except
at very high densities, $v(\rho)$ has the form
$v(\rho)=v_0(1-\rho/\rho_0)$ with a constant $\rho_0$, so that $I_2$
scales like $I_2\propto-v_0\rho^2$.

Although we have set $D_{\mathrm{t}} = 0$ when deriving these results,
it is simple to {establish that} the only direct effect of nonzero
$D_{\mathrm{t}}$ is to add a term $D_{\mathrm{t}}\rho$ to $P_0$
{\cite{Longer}}. There is also an indirect effect on $P_{\mathrm{D}}$
and $P_{\mathrm{I}}$ because $D_{\mathrm{t}}\neq 0$ alters the
correlation functions appearing in $I_1$ and $I_2$.

\section{Numerical methods}\label{methods_section}

All simulation results presented in the main text are obtained for
spherical particles whose centres are confined to two dimensions (the
$xy-$plane) and whose propulsion directions $\vec{u}$ are constrained
to lie in this plane. These particles interact {pairwise through a
  repulsive Weeks-Chandler-Andersen potential:}
\begin{equation}
U(r) = 4\varepsilon \!\left[ \left(\frac{\sigma}{r}\right)^{12} - \left(\frac{\sigma}{r}\right)^{6} \right]\! + \varepsilon 
\label{WCA}
\end{equation}
with an upper cut-off at $r = 2^{1/6}\sigma$, beyond which $U =
0$. Here $\sigma$ denotes the particle diameter, $\varepsilon$
determines the interaction strength, and $r$ is the center-to-center
separation between two particles. The model was studied by solving the
fully overdamped translational and rotational Langevin equations. In
the current section we restore an explicit particle mobility
$v_0/F^\mathrm{a} = \beta D_\mathrm{t}$ rather than setting this to
unity. The Langevin equations then read:
\begin{align}
\dot{\mathbf{r}}_i &= \beta D_{\mathrm{t}} ( \mathbf{F}^{\mathrm{tot}}_i + F^{\mathrm{a}} \vec{u}_i ) + \sqrt{2D_{\mathrm{t}}} \vec{\eta}_i \;, \label{Langevin_t} \\
\dot{\theta}_i &= \sqrt{2D_{\mathrm{r}}} \xi_i \;, \label{Langevin_r} 
\end{align}
where $\mathbf{F}^{\mathrm{tot}}_i$ is the total conservative force
acting on particle $i$, $F^{\mathrm{a}}$ is the constant magnitude of
the self-propulsion force which acts along $\vec{u}_i$,
$D_{\mathrm{t}}$ and $D_{\mathrm{r}} = 3D_{\mathrm{t}}/\sigma^{2}$
denote the translational and rotational diffusivities, respectively;
$\beta = 1/(k_{\mathrm{B}} T)$ is the inverse thermal energy, and
$\vec{\eta}_i(t)$ and $\xi_i(t)$ are zero-mean unit-variance Gaussian
white noise random variables. Simulations were carried out using the
LAMMPS \cite{Plimpton-1995} molecular dynamics package, in a periodic
box with $L_x = L_y = 150 \sigma$ (corresponding to $N \approx 20 000$
particles). The natural simulation units are $\sigma$, $\varepsilon$,
and $\tau_{\mathrm{LJ}} = \sigma^{2}/(\varepsilon \beta
D_{\mathrm{t}})$ for length, energy, and time, respectively. In these
units, a time step of $5 \times 10^{-5}$ was used throughout. As
discussed in \cite{Stenhammar-softmatter} and in Section \ref{Semi}
below, the P\'eclet number Pe $\equiv 3v_0/(D_{\mathrm{r}}\sigma) =
3\beta D_{\mathrm{t}}F^{\mathrm{a}}/(D_{\mathrm{r}}\sigma)$ was varied
by adjusting $D_{\mathrm{r}}$ (and hence $D_{\mathrm{t}}$), keeping a
constant value of $F^{\mathrm{a}} = 24 \varepsilon / \sigma$.

The value of $\rho_0$, the density where the linearly decreasing swim
speed goes to zero, was determined by fitting sampled values at Pe =
40 (\emph{i.e.}, just outside the phase-separated region) of $v(\rho)$
over the density range $[0,1.15]$ to the linear function $v(\rho) =
v_0 (1-\rho/\rho_0)$. The value thus obtained, $\rho_0 \approx 1.19$,
was used in reporting the density data presented in the main text as a
function of $\rho/\rho_0$.

Binodal densities were determined from simulations by coarse-graining
the local density on a grid using a weighting function $w(r) \propto
\exp(-r^{2}_{\mathrm{cut}}/(r^{2}_{\mathrm{cut}} - r^{2}))$, where $r$
is the distance between the particle and a lattice point, and
$r_{\mathrm{cut}}$ is a cut-off distance which was taken to be
slightly larger than the lattice spacing. The local densities thus
obtained were binned and plotted as a probability distribution
function, where the maxima of the two density peaks were taken to
represent the coexistence {densities.}

\section{Semi-Empirical Equation of State}\label{Semi}

We now revert to our convention that the particle mobility is {unity,
  and rewrite Eq.~(14) of the main text as}
\begin{equation}
P_\mathrm{S} = \left(\frac{1}{\textrm{Pe}} + \frac{v(\rho,\textrm{Pe})}{6v_0}{\textrm{Pe}}\right)\sigma \rho v_0\; {.} \label{SEES1}
\end{equation}
Our semi-empirical equation drops the 1/Pe term (which comes from
passive translational diffusion) and assumes that the Pe-dependence in
$v(\rho,\textrm{Pe})$, which arises from Pe-dependence in the bulk
correlators, is negligible.  For $v(\rho)$ we use the fitted linear
function for $v(\rho)$ described above, with the further assumption
that $v = P_{\mathrm{S}} = 0$ for $\rho > \rho_0$ in order to prevent
negative swim speeds (see black curve in Fig.\ 1 of the main text).
With these assumptions (which imply that $\rho_0$ is itself
Pe-independent), the swim pressure scales as $P_{\mathrm{S}} = \sigma
\rho v_0 {\cal{G}}(\rho/\rho_0) \mathrm{Pe}$ with the function
${\cal{G}}(\rho/\rho_0)=v(\rho)/(6v_0)$.  This {ansatz} is confirmed
numerically by comparing datasets with two different Pe in Fig.\ 1 of
the main text.

In the main text we also state the scaling hypothesis
\begin{equation}
P_\mathrm{D} \equiv \sigma\rho v_0{\cal{S}}(\rho/\rho_0,\textrm{Pe}) = \sigma\rho v_0{\cal{S}}(\rho/\rho_0)\; , \label{SEES2}
\end{equation}
The first identity defines a reduced direct pressure ${\cal{S}}$; the
second equality once again requires that Pe has no direct effect on
the correlators (which would enter both through the shape of the
function ${\cal{S}}$ and through $\rho_0$ itself). Again this is
confirmed by comparing $P_{\mathrm{D}}$ for two Pe values in Fig.\ 1
of the main text. Since we choose to vary Pe at fixed $v_0$, a single
$P_{\mathrm{D}}$ function then describes all our simulations; we fit
this as $P_{\mathrm{D}}(\rho) = \alpha(1-\exp(\gamma \rho))$, with
$\alpha$ and $\gamma$ fitting parameters. {Note that $P_\mathrm{D}$ is
  the pressure measured from averaging Eq.~(10) of the main text over
  $\Lambda$ ({see red} curve in Fig.\ 1 of the main text) which is
  mathematically equivalent to using the standard virial relation for
  pairwise additive forces \cite{Allen,Doi}.}

The above scaling forms (\ref{SEES1}) and (\ref{SEES2}) assume that,
once pressures are non-dimensionalized by a factor $\sigma \rho v_0$
(recalling that the mobility is unity), there can be no further
dependence on $v_0$ except via the dimensionless combination Pe. This
is true for hard particles, but could fail for softened interactions
as actually used in our simulations: in particular, at large $v_0$ the
effective diameter of the particles seen in collisions will be less
than $\sigma$; see \cite{Stenhammar-softmatter}.  Accordingly the best
route for testing the scalings with Pe is to vary this at fixed $v_0$,
as we do here.

\begin{figure}
\includegraphics[width=0.35\textwidth]{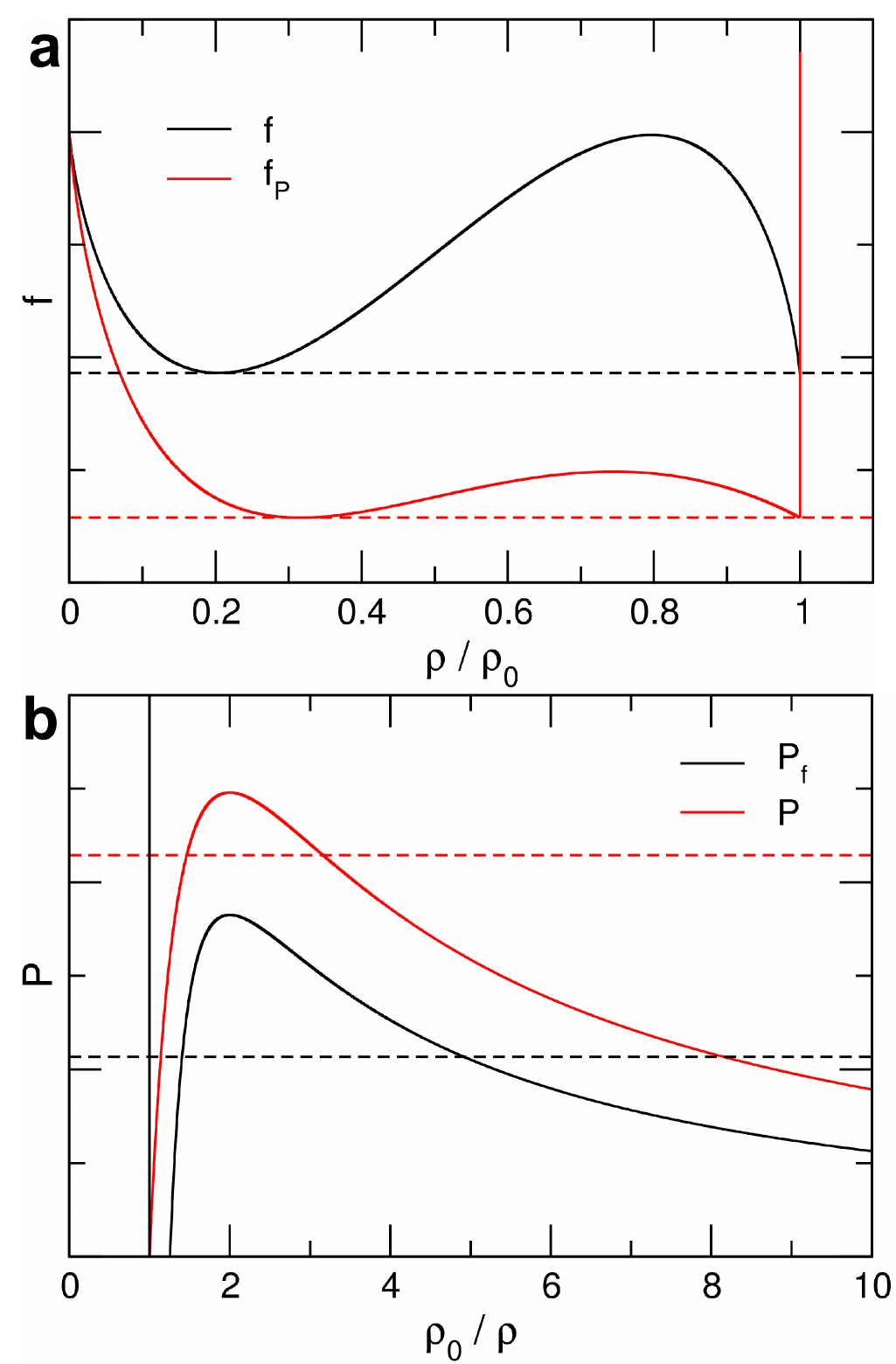}
\caption{(a) Upper curve: {CTC} (dashed) {on $f$ (solid)}  based on (\ref{sup1}). Lower curve: {CTC} (dashed) on {$f_{P}$ (solid)} based on (\ref{sup4}). 
In each case a linear {term has} been subtracted to make the common tangent horizontal. (b)  
\label{fig:sup1}
Upper curve: {MC} (dashed) on the mechanical pressure $P =
P_{\mathrm{S}}+P_{\mathrm{D}}$ (solid) based on (\ref{sup2}). Lower
curve: {MC} (dashed) on the pressure $P_{f}$ (solid) based on
(\ref{sup3}). The curves in (a) and (b) are rescaled/displaced
vertically for improved visibility.}
\end{figure}

\section{Constructions of the binodals}

As defined in the main text, we consider four routes (in two
equivalent pairs) to calculate binodal densities in the high-Pe
limit. We use `thermodynamic' routes (via $f,P_f$) and {`mechanical'}
routes (via $P,f_P$), relying on {the Maxwell equal-area construction
  (MC) and common tangent construction (CTC)} as appropriate.

Method 1 starts from the effective free energy of
\cite{TC,CT} 
\begin{equation}
\tilde f(\rho) = k_{\mathrm{B}}T\left(\rho(\ln\rho-1) + \! \int_0^\rho \!\!\!\ln v(u) \,\dif u\right) ,
\end{equation}
where for ABPs $v = v_0(1-\rho/\rho_0)$. This we supplement by a
hard-core cutoff at $\rho = \rho_0$; {hence $f$ obeys}
\begin{equation}
f = \tilde f \quad\hbox{\rm for}\quad \rho \le \rho_0 \;,\quad\hbox{\rm else}\quad f = +\infty \;.
\label{sup1}
\end{equation}
The {CTC} is then performed on $f$ (see Fig.\ \ref{fig:sup1}a).
Method 2 starts from the mechanical pressure $P = P_{\mathrm{S}} +
P_{\mathrm{D}}$, representing $P_{\mathrm{D}}$ as a hard-core cutoff:
$P_{\mathrm{D}} = 0$ for $\rho \le \rho_0$ and $P_{\mathrm{D}} =
+\infty$ for $\rho>\rho_0$. $P$ therefore obeys
\begin{equation}
P = \frac{\rho v_0^2}{2D_{\mathrm{r}}}(1-\rho/\rho_0) \;\,\hbox{\rm for}\;\, \rho \le \rho_0 \;,\;\,\hbox{\rm else}\;\, P = +\infty \;. \label{sup2}
\end{equation}
{The MC} is then applied to $P$ (see Fig.\ \ref{fig:sup1}b). 
Method 3 constructs the thermodynamic pressure $P_{f} = \rho f'-\rho$ from $f$, that is
\begin{equation}
P_{f} = \rho\tilde f' - \tilde f \quad\hbox{\rm for}\quad \rho \le \rho_0 \;,\quad\hbox{\rm else}\quad P_{f} = +\infty \label{sup3}
\end{equation}
and then applies the {MC} to $P_f$. By mathematical necessity, this
gives the same binodals as Method {1} (see Fig.\ \ref{fig:sup1}b).
Method 4 constructs a different effective free energy $f_{P}$ such
that $P = \rho f_{P}'-f_{P}$.  The {result is}
\begin{equation}
\begin{split}
f_{P} = \frac{\rho v_0^2}{2D_{\mathrm{r}}}\!\left[\rho(\ln \rho -1)-\frac{\rho^2}{\rho_0}\right]\! \quad\hbox{\rm for}\quad \rho \le \rho_0 \;,& \\
\;\hbox{\rm else}\quad f_{P} = +\infty &
\end{split}\label{sup4}
\end{equation}
from which binodals are found by the {CTC} on $f_P$. By mathematical
necessity, this gives the same binodals as Method {2} (see Fig.\
\ref{fig:sup1}a).

As is clear from Fig.\ \ref{fig:sup1}, Method 1 (or 3) based on $f$
(or $P_{f}$) gives different binodals from Method 2 (or 4) based on
$P$ (or $f_{P}$). These calculations all use the sharp cut-off
approximation and hence the resulting binodals refer to the asymptotic
limit of high Pe only. In this limit, Method 1 (or 3) is clearly more
accurate than Method 2 (or 4) (see Fig.\ 2 of the main text).

However, we do not know how to generalize Method 1 (or 3) to the case
of finite Pe, since we lack a theory for constructing the direct
interaction contributions to $f$ or $P_{f}$. Method 2 (or 4) does
generalize, allowing use of the semi-empirical expressions for
$P_{\mathrm{S}}$ and $P_{\mathrm{D}}$ described above and in the main
text. However, as shown there (see Fig.\ 2 of the main text) the
results are unsatisfactory.

None of these methods allows for nonlocal contributions, which are
shown in \cite{ModelB} to alter the common tangent construction found
by Method 1. Similar nonlocal terms are also known to arise in
calculations of mechanical force balance at phase coexistence in
systems undergoing continuous driving, such as in shear banding
\cite{olmsted}; they are likewise unjustifiably neglected by Method 2
(or 4). We conclude, as stated in the main text, that no adequate
theory of phase equilibria in ABPs yet exists.


\begin{thebibliography}{99}
\bibitem{footroll} Any self-propelled entity whose motility depends on frictional contact with a support (such as human walking, cell crawling \cite{elsen}, vibrated granular materials \cite{menon}, or colloids that move by rolling on a surface \cite{rollers}) is exchanging momentum with an external reservoir (the support).
\bibitem{TI} We assume, without loss of generality, that translational invariance in $y$ is maintained even if the system undergoes phase separation into two or more isotropic phases.
\bibitem{LDF} The large deviation functional (LDF, or effective free
  energy) $\mathcal{F}[\hat\rho(\vec{r})]/(Vk_\mathrm{B}T)$ for the fluctuating
  density $\hat\rho$ in a nonequilibrium system is defined as
  $-\ln(\mathrm{Pr}[\hat\rho (\vec{r})])/V$ where Pr is the
  steady-state probability distribution \cite{Ellis}. In \cite{TC,CT},
  it is shown that $\int f(\hat\rho)\dif^dr/(Vk_\mathrm{B}T)$ is, within the
  local approximation, the LDF for a system of particles with a
  density-dependent swim speed {$v(\rho)$.}
\bibitem{softfoot} An additional simulation at Pe = 500 gave a lower binodal
  value, $\rho/\rho_0 \simeq 0.08$. This may be due to the
  non-local gradient terms identified in \cite{ModelB}.
\end{thebibliography}
\end{document}